\documentclass{aastex63}

\usepackage{amsmath}

\usepackage{xcolor}
\newcommand{\COMON}{\begin{color}{blue}}
\newcommand{\COMOFF}{\end{color}}

\submitjournal{ApJ}

\shorttitle{Stream interaction effects on ICME propagation}
\shortauthors{Winslow et al.}

\begin{document}

\title{The effect of stream interaction regions on ICME structures
observed in longitudinal conjunction}

\correspondingauthor{Reka M. Winslow}
\email{reka.winslow@unh.edu}

\author[0000-0002-9276-9487]{Reka M. Winslow}
\affiliation{Institute for the Study of Earth, Oceans, and Space, University of New Hampshire, Durham, NH, USA}

\author[0000-0002-5681-0526]{Camilla Scolini}
\affiliation{Institute for the Study of Earth, Oceans, and Space, University of New Hampshire, Durham, NH, USA}
\affiliation{University Corporation for Atmospheric Research, Boulder, CO, USA}

\author[0000-0002-1890-6156]{No\'e Lugaz}
\affiliation{Institute for the Study of Earth, Oceans, and Space, University of New Hampshire, Durham, NH, USA}

\author{Antoinette B. Galvin}
\affiliation{Institute for the Study of Earth, Oceans, and Space, University of New Hampshire, Durham, NH, USA}



\begin{abstract}
We study two interplanetary coronal mass ejections (ICMEs) observed at Mercury and 1~AU by spacecraft in longitudinal conjunction, investigating the question: what causes the drastic alterations observed in some ICMEs during propagation, while other ICMEs remain relatively unchanged? Of the two ICMEs, the first one propagated relatively self-similarly, while the second one underwent significant changes in its properties. We focus on the presence or absence of large-scale corotating structures in the ICME propagation space between Mercury and 1~AU, that have been shown to influence the orientation of ICME magnetic structures and the properties of ICME sheaths.
We determine the flux rope orientation at the two locations using force-free flux rope fits as well as the classification by \citet{Nieves2019}. We also use measurements of plasma properties at 1~AU, the size evolution of the sheaths and ME with heliocentric distance, and identification of structures in the propagation space based on in situ data, remote-sensing observations, and simulations of the steady-state solar wind, to complement our analysis.
Results indicate that the changes observed in one ICME were likely caused by a stream interaction region, while the ICME exhibiting little change did not interact with any transients between Mercury and 1~AU. This work provides an example of how interactions with corotating structures in the solar wind can induce fundamental changes in ICMEs. Our findings can help lay the foundation for improved predictions of ICME properties at 1~AU.
\end{abstract}



\section{Introduction} 

Coronal mass ejections (CMEs) are large eruptions of plasma and magnetic field from the Sun \citep[e.g.][]{Webb2012},
observed to occur at a typical rate of 0.5--5 CMEs per day depending on the phase of the solar cycle \citep[e.g.][]{Lamy2019}. 
They are observed by coronagraphs \citep{Illing1985, Vourlidas2013} as they propagate away from the Sun into interplanetary space, and are eventually probed in situ by spacecraft monitoring the conditions of the interplanetary medium, where they are termed interplanetary CMEs \citep[ICMEs; e.g.][]{Zurbuchen2006, Kilpua2017}.
CMEs consist of magnetically-dominated (i.e.\ low-$\beta$) plasma that can be often described by a magnetic flux rope morphology, i.e.\ by a helical magnetic field wrapping around a central axis. 
At 1~AU, ICMEs often present as an interplanetary shock followed by a sheath region of compressed solar wind material,{ and followed by the magnetic ejecta (ME) \citep[e.g.][]{Richardson2010, Kilpua2017, Jian2018}, i.e., the magnetically dominated substructure.} MEs are observed passing over Earth at an average rate of 1--2 per month \citep{Richardson2010}, where together with their driven shocks and sheaths they are known to be one of the main causes of strong geomagnetic disturbances \citep[e.g.][]{Gosling1991, Zhang2007, Lugaz2016, Kilpua2017, Kilpua2019}. In more recent years, studies of ICMEs as key space weather drivers throughout the heliosphere have shed light on their impact on planetary environments at heliocentric distances other than Earth, where the ICME characteristics are intrinsically different than at 1~AU \citep[e.g.][]{Burlaga1982, Liu2005, Ebert2009, Winslow2015, Good2016, Lee2017}.

ICME properties {are known to} evolve during propagation in interplanetary space as they expand and interact with the solar wind and other transients therein \citep[e.g.][]{Manchester2017}.
The large-scale evolution of ICMEs propagating through interplanetary space is primarily shaped by two effects: the expansion of the ME, which controls its internal density, pressure, magnetic field magnitude and size \citep[e.g.][]{Demoulin2009}; and the interaction with the surrounding solar wind, which controls the ICME kinematic properties, and is often described in terms of a drag force \citep{Cargill2004, Vrsnak2010}. As a result, ICME shocks and sheath regions are also the result of the interplay between ICME propagation and expansion \citep{Kilpua2017}.
In this respect, ICMEs of differing coronagraph speeds have been observed to have different responses to interaction with the solar wind. It is thought that fast ICMEs go through a period of impulsive acceleration followed by rapid deceleration and finally propagation at nearly constant speed, while slow ICMEs catch up to the solar wind speed \citep{Liu2013}.

Through observational and modeling work, studies have shown that during propagation ICMEs undergo a number of changes, particularly as a consequence of the interaction with high-speed streams (HSSs), stream/corotating interaction regions (SIRs/CIRs), and the heliospheric current/plasma sheet (HCS/HPS)
\citep[e.g.][]{Odstrcil1999b, Odstrcil1999c, Rodriguez2016, Winslow2016, Zhou2017, Heinemann2019, Liu2019}.
Effects include kinks and deformations of the ICME flux rope magnetic structure \citep{Manchester2004, Savani2011, Isavnin2016}, large-scale deformations of the ICME front convexity \citep{Odstrcil1999, Riley2004}, as well as local magnetic field distortions \citep{Torok2018}.
Furthermore, erosion of their internal magnetic flux may occur as a consequence of magnetic reconnection  with the surrounding solar wind \citep{Dasso2006, Lavraud2014, Ruffenach2015, Pal2020}, in turn affecting the ICME size.
ICMEs may also get deflected \citep{Kay2013, Kay2015, Wang2014, Zhuang2019} and rotated \citep{Yurchyshyn2008, Lynch2009, Kliem2012, Isavnin2014}.
Finally, interaction with other ICMEs can also strongly contribute to alter their kinematics, magnetic properties, and potential space weather impact during propagation \citep[e.g.][]{Liu2014, Lugaz2017, Shen2017, Scolini2020}.

The properties of ICME shocks and sheaths also evolve during propagation in interplanetary space \citep{Janvier2019, Lugaz2020a, Lugaz2020b, Salman2020}. Depending on the driver's and downstream solar wind's characteristics, their evolution may result in enhanced magnetic field amplitudes, increased magnetic turbulence, and the formation of planar magnetic structures, all of which are likely to increase the potential space weather impact of such structures with heliocentric distance \citep{Lugaz2016, Palmerio2016, Kilpua2017, Good2020}.

Direct observational studies of the radial evolution of ICMEs are intrinsically linked to our capability to perform high-quality in situ measurements of individual events via multi-spacecraft crossings through approximately the same portion of the ICME structure at different heliocentric distances. Such studies are generally constrained by the combination of a number of factors, including: the limited number of assets available (in terms of missions and instruments); the requirement of continuous observations of the solar wind properties over a time window of hours to days, which are typically difficult to achieve in the case of planetary missions; and the need for appropriate spacecraft trajectories through the ICME structure, necessary to measure some of the ICME properties (e.g. their orientation) with a high degree of confidence.
To overcome the scarcity of data, statistical studies have been proven to be extremely powerful in characterising the overall trends affecting ICME evolution, most importantly in the case of ICMEs observed by multiple spacecraft in longitudinal conjunctions \citep{Good2019, Salman2020}. However, these kinds of studies only provide an average picture of the actual evolution of individual ICMEs, and -- perhaps more importantly in the context of this work -- they cannot dive deep into the analysis of individual events to determine the causes behind non-ideal evolutionary behaviour \citep[recently reported by][]{Lugaz2020a}. 

For this reason, studies focusing on individual ICME events observed by multiple spacecraft during periods of longitudinal conjunction are extremely insightful for our understanding of the various phenomena controlling the evolution of ICME magnetic structures.
On the one hand, in situ studies by \citet{Good2015, Good2018} of two ICMEs observed in perfect conjunction at Mercury and STEREO-B have showcased two events where the evolution of the large-scale flux rope magnetic field within the ICMEs remained essentially self-similar during propagation to 1~AU, also in agreement with previous studies of longitudinal conjunction events in the outer heliosphere \citep[e.g.][]{Nakwacki2011}.
Similarities in the magnetic flux rope observations at different heliocentric distances had also been previously reported by \citet{Moestl2012} at Venus Express and STEREO-B, despite the $\sim 18^\circ$ longitudinal separation between the spacecraft. 
On the other hand, not all ICMEs are observed to evolve self-similarly in interplanetary space.
For example, an event study by \citet{Nieves2012} using in situ observations from MESSENGER and Wind showed the first direct evidence of a significant re-orientation of the ICME flux rope axis during propagation through the inner heliosphere, although there was $\sim 20^\circ$ longitudinal separation between the spacecraft.
In \citet{Winslow2016}, we also showcased an ICME flux rope observed in near-perfect longitudinal conjunction, which underwent a significant increase in ICME complexity as it propagated from Mercury to 1~AU due to interaction/reconnection with the HCS/HPS. 
More recently, \citet{Good2019} and \citet{Lugaz2020a} reached somewhat opposite conclusions regarding the typical evolutionary behaviour of ICMEs (i.e. ideal vs non-ideal), based on the results of statistical analyses carried out over different sets of ICMEs observed in longitudinal conjunction.

The varied results from previous studies therefore raise the question: what causes some ICME structures, such as the flux rope and sheath, to change drastically during propagation, while in other ICMEs these stay relatively self-similar? More generally, what are the main drivers of increases in ICME complexity {during propagation in the heliosphere}? These past works and the considerations above highlight the need for more in-depth studies of ICME global changes from the innermost heliosphere to 1~AU, in order to determine the causes of drastic alterations in ICME structures during propagation. 
 
In this work, we tackle the questions above by conducting in-depth analyses of two ICME case studies, using in situ measurements at MESSENGER and spacecraft located near 1~AU to investigate changes in the global ICME structure that may have occurred during propagation. We also use these observations to identify any intervening ICME, SIR/CIR and HCS/HPS that might have affected the propagation of the ICMEs of interest. In addition, to place any observed changes in ICME properties during propagation in context, we simulate the background solar wind using the WSA-ENLIL model available for runs on request at the NASA Community Coordinated Modeling Center\footnote{\url{https://ccmc.gsfc.nasa.gov}} (CCMC), which enables improved identification of solar wind structures in the propagation space of the ICMEs under study.

\medskip
This paper is structured as follows.
In Section~\ref{sec:methods}, we provide an overview of the methods used to investigate in situ plasma and magnetic field observations of ICMEs and their driven sheaths at different spacecraft and heliocentric distances.
In Section~\ref{sec:event1}, we present one case study ICME observed in almost perfect longitudinal conjunction at MESSENGER and ACE, and which exhibited little change in the magnetic and plasma properties during its propagation from Mercury's to Earth's orbit. 
On the other hand, in Section~\ref{sec:event2}, we present a second case study ICME that, although having been observed by MESSENGER and STEREO-B at longitudinal separation of $\sim 30^\circ$, exhibited significant changes in it is internal and sheath properties between Mercury and 1~AU that cannot be attributed to the longitudinal separation alone. We further compare the two cases and investigate the possible causes of the alterations observed in the second event.
Finally, in Section~\ref{sec:conclusions}, we discuss the main findings and present our conclusions.

\section{Methods: Flux rope orientation and plasma parameters}
\label{sec:methods}


To investigate the ICME properties at 1~AU, we use in situ magnetic field and plasma data from the
magnetometer \citep[MAG;][]{Smith1998} and the Solar Wind Electron, Proton and Alpha Monitor \citep[SWEPAM;][]{McComas1998}
on board the Advanced Composition Explorer (ACE) mission \citep{Stone1998}, orbiting the Sun--Earth Lagrange~1 (L1) point;
and in situ magnetic field and plasma data from the
magnetometer \citep[MAG;][]{Acuna2008},
the In situ Measurements of Particles And CME Transients \citep[IMPACT;][]{Luhmann2008}, 
and the Plasma And Suprathermal Ion Composition \citep[PLASTIC;][]{Galvin2008}
on board the Solar TErrestrial RElations Observatory -- Behind (STEREO-B) \citep{Kaiser2008} spacecraft.
At Mercury, we make use of in situ data provided by the magnetometer \citep[MAG;][]{Anderson2007} on board the MErcury Surface, Space ENvironment, GEochemistry, and Ranging \citep[MESSENGER;][]{Solomon2007} mission.
Remote-sensing CME observations from Earth's viewpoint are provided by the Large Angle and Spectrometric Coronagraph \citep[LASCO;][]{Brueckner1995} on board the Solar and Heliospheric Observatory \citep[SOHO;][]{Domingo1995} satellite, while observations from additional vantage points are provided by the Sun Earth Connection Coronal and Heliospheric Investigation \citep[SECCHI;][]{Howard2008} instrument on board STEREO-B and STEREO -- Ahead (STEREO-A).

With regards to MESSENGER data, we note that MESSENGER provided magnetic field data; however, it did not provide reliable plasma measurements in the solar wind. 
The spacecraft was in orbit around Mercury during the two ICME events highlighted in this paper. We thus used the methods described in \citet{Winslow2013} to identify the crossings of Mercury's bow shock and magnetopause (and therefore the magnetosphere) and remove them from the data analyzed and presented here. Both case studies presented below, but especially the first one, have numerous bow shock crossings throughout the ICME passage as a result of the motion of the bow shock relative to the spacecraft. These cause MESSENGER to remain in the planetary magnetosheath for extended periods where the interplanetary magnetic field (IMF) is modified by Mercury's bow shock. This further complicates studying the ICME characteristics because during some periods there is little pristine IMF data available. For these two case studies, we started by identifying every instance of bow shock crossing during the ICME passage and adding in all possible times when the spacecraft was outside of the magnetosheath. To further increase the useable MESSENGER ICME data, we followed the method of \citet{Lugaz2020b} to extrapolate the IMF measurements by dividing the measured magnetic field inside the magnetosheath by the jump experienced through the bow shock for the magnitude and each individual component. We refer the reader to \citet{Lugaz2020b} for further details, and we note that this procedure allowed us to recover valuable additional hours of magnetic field data during both ICME passages described below.

At MESSENGER, due to the lack of direct solar wind observations \citep{Andrews2007} we also developed a method to estimate the solar wind dynamic pressure using magnetic field measurements (see \citet{Winslow2017} for details).  Using the modified Newtonian approximation during magnetospheric transits of Mercury by MESSENGER (described in detail in Section~2.2 and Equation~(4) of \citet{Winslow2017}), we determined a proxy for the solar wind ram pressure from the magnetopause magnetic field strength. This method has been tested in both \citet{Winslow2020} and \citet{Lugaz2020b} and has been found to yield results that compare favorably with measured dynamic pressure values at 1~AU scaled back to Mercury's heliocentric distance.

To investigate changes in the global magnetic field structure of ICMEs as detected in situ at various heliocentric distances, 
we employ two different methods: 
(1) a constant-$\alpha$ force-free field fitting \citep{Burlaga1988, Lepping1990}; 
and (2) the classification method proposed by \citet{Nieves2019}, which allows the classification of ICME flux rope structures based on the extent and number of observed magnetic field rotations through the analysis of magnetic hodograms. This method avoids the introduction of further assumptions on the 3D magnetic configuration as required by more advanced fitting techniques, and it hence appears particularly suitable to investigate and compare the ICME magnetic complexity measured at various heliocentric distances and spacecraft.
To investigate and reconstruct the global structures of the ICME magnetic field and determine their changes during propagation, we apply the two aforementioned techniques to magnetic field data at both Mercury and at 1~AU from single spacecraft crossings through the MEs. The comparison of the reconstructed global properties obtained at the two heliocentric distances then allows us to determine if substantial changes in the flux rope have occurred during propagation. 
We note that because of the lack of direct plasma density measurements at Mercury, some more sophisticated flux rope reconstruction techniques, such as the Grad-Shafranov method \citep{Hau1999} or the elliptical cross-section technique \citep{Hidalgo2002}, which require information on the plasma velocity and thermal pressure, are essentially impossible to apply.


To characterize the solar wind conditions along and around the propagation path of the ICMEs under study, we {use heliospheric simulations obtained from} the WSA-ENLIL model {and} available on the {NASA} CCMC website.
ENLIL \citep{Odstrcil2003} is a time-dependent three-dimensional (3-D) magnetohydrodynamic (MHD) model of the inner heliosphere designed to model the ambient solar wind as well as CMEs propagating through it. 
In this work, we use results obtained with ENLIL version 2.8f in combination with the Wang-Sheeley-Arge (WSA) semi-empirical coronal model \citep{Arge2000, Arge2004}. 
{In this work, we performed simulations using} a heliospheric computational domain spanning between 0.1~AU and Mars orbit in the radial direction, and covering $\pm 60^\circ$ in latitude and $\pm 180^\circ$ in longitude in Heliocentric Earth EQuatorial (HEEQ). In the simulations, a medium resolution with 512 grid cells in the radial direction, 60 grid cells in the latitudinal direction, and 180 grid cells in the longitudinal direction, was used. As input for the WSA coronal model, {we used daily-updated synoptic} magnetograms generated by the National Solar Observatory (NSO) {Global Oscillation Network Group (GONG)}. {The CMEs were modeled using the cone model, and initialized with parameters taken from the Space Weather Database Of Notifications, Knowledge, Information (DONKI\footnote{\url{https://kauai.ccmc.gsfc.nasa.gov/DONKI/search/}}) catalog.}
{In addition to performing new simulations of the solar wind and CME conditions, we also considered pre-existing simulations of the steady-state ambient solar wind alone, in order to better contextualize the environment through which the CMEs propagated.}

In what follows, we describe in detail two CMEs and their propagation between Mercury and 1~AU, with one event affected, and the other not affected, by corotating structures in the propagation space. 

\section{CME with no corotating structures in the propagation space}
\label{sec:event1}

The first case study considered is a CME launched from the Sun on 9 July 2013 (first observed in LASCO~C2 at 15:12~UT) that impacted both Mercury and Earth, which were in radial alignment at this time (within 3$^{\circ}$ of longitudinal separation). This event had been previously studied by \citet{Moestl2018} to test a new space weather modeling tool, and by \citet{Lugaz2020b} to investigate its evolution and sheath region. Given the detailed analysis of this CME/ICME documented in both \citet{Moestl2018} and \citet{Lugaz2020b}, here we only provide a summary of the remote-sensing and in situ observations in the corona and heliosphere, and rather focus on the investigation of in situ properties that were not addressed in those papers. We refer the reader to the aforementioned papers for more detailed information on the remote observations and kinematics of this CME.

At the Sun, this CME was associated with {an obvious} filament eruption, occurring around 14~UT at $\theta\sim15^\circ$, $\phi\sim-5^\circ$ in Stonyhurst coordinates on the solar disk.
{Prior to the eruption, the filament exhibited a clear inverse-S topology \citep[see Figure~2 in][]{Moestl2018}, indicative of magnetic fields characterized by a left-handed (negative) chirality \citep[e.g.][]{Green2007, Green2018}.
Additionally, measurements of the photospheric magnetic field suggest the axial field of the magnetic flux rope formed in the source region was inclined by roughly $45^\circ$ with respect to the solar equator \citep[as marked by the orientation of the magnetic polarity inversion line; e.g.][]{Titov1999}. The magnetic polarity signs at two sides of the polarity inversion line further suggests the axial field was pointing towards the south-west direction \citep[as shown in Figure~2 in][]{Moestl2018}. 
The resulting magnetic flux rope is therefore expected to be a mid-inclination configuration between NWS and WSE types (following the classifications by \citet{Bothmer1998} and \citet{Mulligan1998}) {\citep[as indicated in Figure~2 in][]{Moestl2018}}.
}

The CME was later observed in the corona by the LASCO~C2 and C3 instruments (starting at 15:12~UT and 16:30~UT, respectively) on board SOHO, as well as by the COR2 instruments on board STEREO-A and STEREO-B (first observations at 15:24~UT and 15:36~UT, respectively).
Previous studies by \citet{Hess2017} and \citet{Moestl2018} using the Graduated Cylindrical Shell model \citep[GCS;][]{Thernisien2006, Thernisien2009} estimate a CME de-projected speed in the corona around $550-600$~km/s, with a main direction of propagation towards $\theta\sim-1^\circ$, $\phi\sim1^\circ$ in HEEQ coordinates, i.e.\ close to the Sun--Mercury--Earth line. 
{An overview of the white-light CME observations in the corona, including a visualisation of the CME 3D reconstruction based on the GCS model, can be found in Figure~3 in \citet{Moestl2018}.}
By comparing the source region location with the CME direction of propagation in the corona, \citet{Moestl2018} concluded that a deflection of about $20^\circ$ away from the initial direction took place during the early CME evolution.
Coronal and heliospheric observations in the days before and after the CME eruption also show little front-sided solar activity, indicating this event likely propagated without interacting with any other CMEs from the Sun to the Earth.


\begin{figure}
\centering
{\includegraphics[width=\hsize]{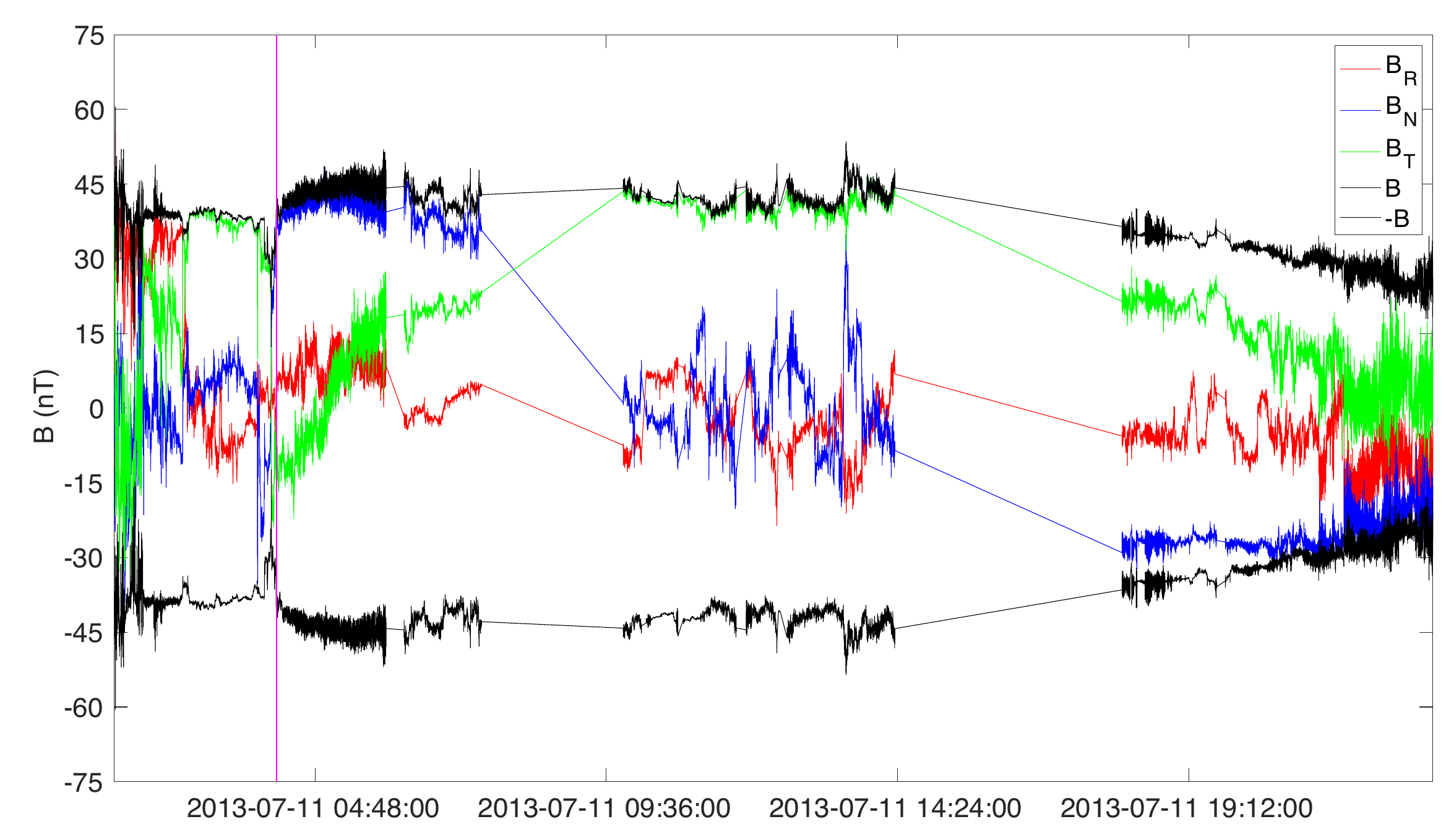}}
\caption{
Magnetic field data for the July 9, 2013 CME event, measured in situ at MESSENGER on July 11, 2013. The data are in RTN coordinates with $B_R$ in red, $B_T$ in green, $B_N$ in blue. The magnetic field magnitude $B$ and its negative $-B$ are in black. The same color coding for the magnetic field data is used in the following figures as well. The ICME shock arrived at Mercury while MESSENGER was still inside the magnetosphere; therefore, we only show data starting at 1:30 UT (i.e. 25 mins past the shock arrival) after MESSENGER had exited the magnetosphere, and the data interval shown goes until the end of the ICME on July 11 at 23:14~UT (after which MESSENGER re-entered the magnetosphere). The vertical magenta line denotes the start time of the ME. The data gaps correspond to MESSENGER's passage through Mercury's magnetosphere. 
} 
\label{fig:event1_messenger}
\end{figure}

\begin{figure}
\centering
{\includegraphics[width=\hsize]{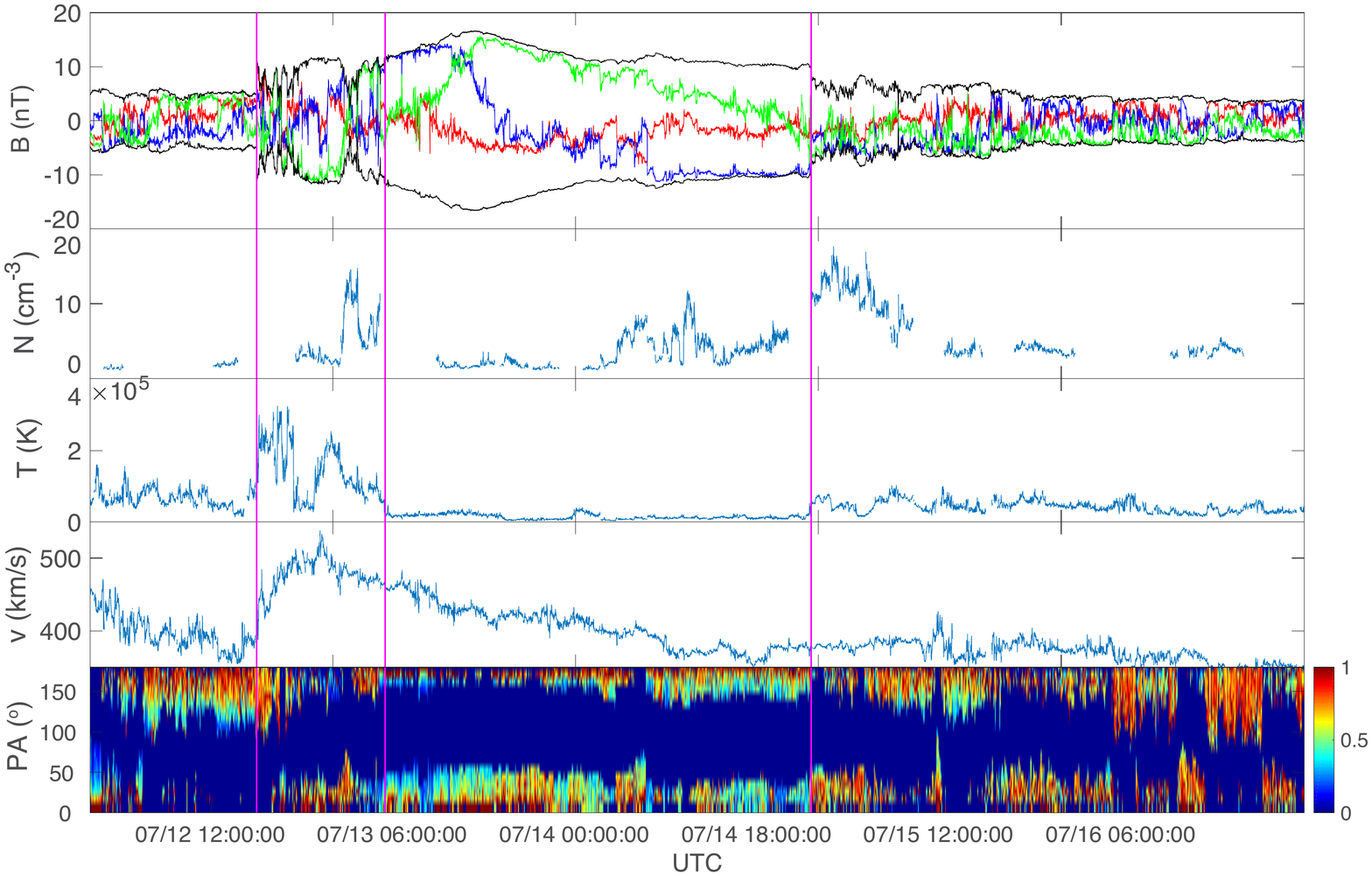}}
\caption{
Data for the July 9, 2013 CME event measured in situ at ACE on July 12-14, 2013. The first panel shows the magnetic field data in RTN coordinates, using the same colors as in Figure~\ref{fig:event1_messenger}.
The second, third, and fourth panels show the proton number density, temperature, and speed, respectively.
The last panel shows the normalized suprathermal electron pitch angle distribution. The vertical magenta lines denote the start times of the shock and ME, and the end time of the ME.
} 
\label{fig:event1_ace}
\end{figure}

ICME signatures were later detected at MESSENGER (Figure~\ref{fig:event1_messenger}) and ACE (Figure~\ref{fig:event1_ace}) about two and three days after the eruption, respectively.
At the time of the ICME impact, the MESSENGER spacecraft was orbiting Mercury, which was located at HEEQ coordinates $(r,\theta,\phi) = (0.45~\mathrm{AU}, -2^\circ, 3^\circ)$,
while ACE was orbiting L1 at HEEQ coordinates $(r,\theta,\phi) = (1.02~\mathrm{AU}, 4^\circ, 0^\circ)$.
The ICME shock arrived at MESSENGER on July 11 (day of year 192) at 01:05~UT, while the ME started on July 11 at 04:10~UT and ended at 23:14~UT on the same day. These boundary choices lead to a sheath length of $\sim 3.1$~hrs and an ME length of $\sim 21.3$~hrs, corresponding to a sheath-to-ME duration ratio equal to 0.16.
The ICME shock arrived at ACE on July 12 at 16:28~UT, while the magnetic ejecta arrived on July 13 at 05:10~UT and ended at 23:18~UT on July 14. At 1~AU, the ICME was therefore observed to have a sheath length of $\sim 12.7$~hrs and an ME length of $\sim 42.1$~hrs, corresponding to a sheath-to-ME duration ratio equal to 0.3.
At both spacecraft, the sheath-to-ME duration ratio is lower than typical values reported by \citet{Janvier2019}, 
and an investigation of the sheath and ME individual duration reveals this is due to the fact that the slightly longer-than-typical sheath was followed by a very extended ejecta compared to typical observations at both heliocentric distances \citep{Lugaz2020b}.

We note that at MESSENGER we have chosen 04:10~UT as the start time of the ME in order to be consistent with the flux rope start time, i.e.\ when smooth rotations of the magnetic field components start in Figure~\ref{fig:event1_messenger}. This choice is also consistent with empirical arguments on the expansion/accumulation of material into the sheath during propagation based on typical observations of the sheath increase factor between Mercury and 1~AU \citep{Janvier2019}. In fact, with these choices of boundaries, the sheath and ME increase by factors of 4.1 and 2, respectively, between Mercury and 1~AU. For comparison, typical factors of increase in sheath and ME size of 5 and 2.7, respectively, have been reported by \citet{Janvier2019} based on observations of statistical sets of ICMEs at Mercury's ($\approx 0.4$~AU) and Earth's ($1$~AU) orbits.
%

For this ICME event, we showed in \citet{Lugaz2020b} that the estimates of the solar wind dynamic pressure at MESSENGER using the methodology described in Section~\ref{sec:methods} are well matched by the values obtained from the scaling of 1~AU $P_{dyn}$ data using a $1/r^2$ dependence. In situ observations therefore suggest a fairly typical decay of the sheath dynamic pressure between Mercury and 1~AU.
The suprathermal electron pitch angle distribution data at ACE completes the picture of the ICME at 1~AU, showing clear bidirectional electrons throughout the large majority of the ME period. 
Counter-streaming electrons inside ICMEs are interpreted as signatures of the passage of a magnetic structure still connected at both ends to the Sun \citep[e.g.][]{Zurbuchen2006, Kilpua2017}, thus these measurements at ACE imply that the ICME did not go through significant reconnection during its propagation to 1 AU.

We previously showed in \citet{Lugaz2020b} that the magnetic field measurements at Mercury matched those at 1~AU remarkably well after scaling the field to account for CME expansion during propagation. However, in that work, the ME orientation and complexity were not analyzed in detail. Using the two different magnetic field characterization and modeling methods described in Section~\ref{sec:methods}, we also confirm below that the global magnetic field structure did not change significantly between Mercury and 1~AU.
\begin{figure}
\centering
{\includegraphics[width=\linewidth]{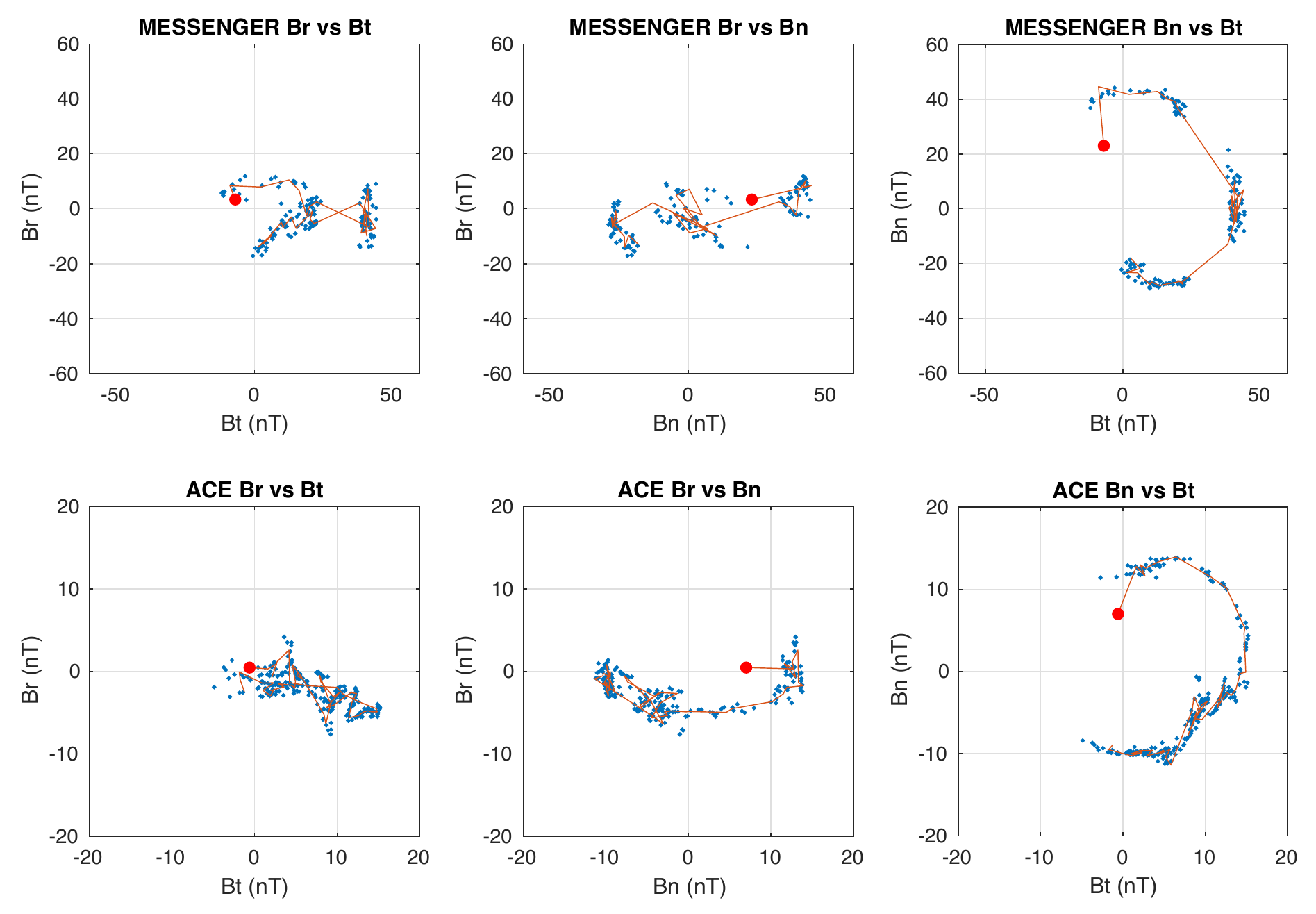}} 
\caption{Magnetic hodograms of the July 9, 2013 CME event at MESSENGER and ACE, in RTN coordinates. The red dots mark the initial values of the magnetic field components.}
\label{fig:event1_hodograms}
\end{figure}

In particular, following the classification of \citet{Nieves2019}, we find that the magnetic ejecta is compatible with a NWS left-handed flux rope with orientation $\phi\sim 90^{\circ}$ and $\theta\sim 0^{\circ}$ (in Radial-Tangential-Normal, hereafter RTN, coordinates) at both spacecraft, {in agreement with the photospheric magnetic field measurements and EUV observations of the source region discussed above.} From the magnetic hodograms shown in Figure~\ref{fig:event1_hodograms}, we find that the magnetic ejecta looks like an $F^+$ flux rope at both spacecraft, i.e.\ with no change in classification between the two locations.

\begin{figure}
\centering
{\includegraphics[width=\linewidth]{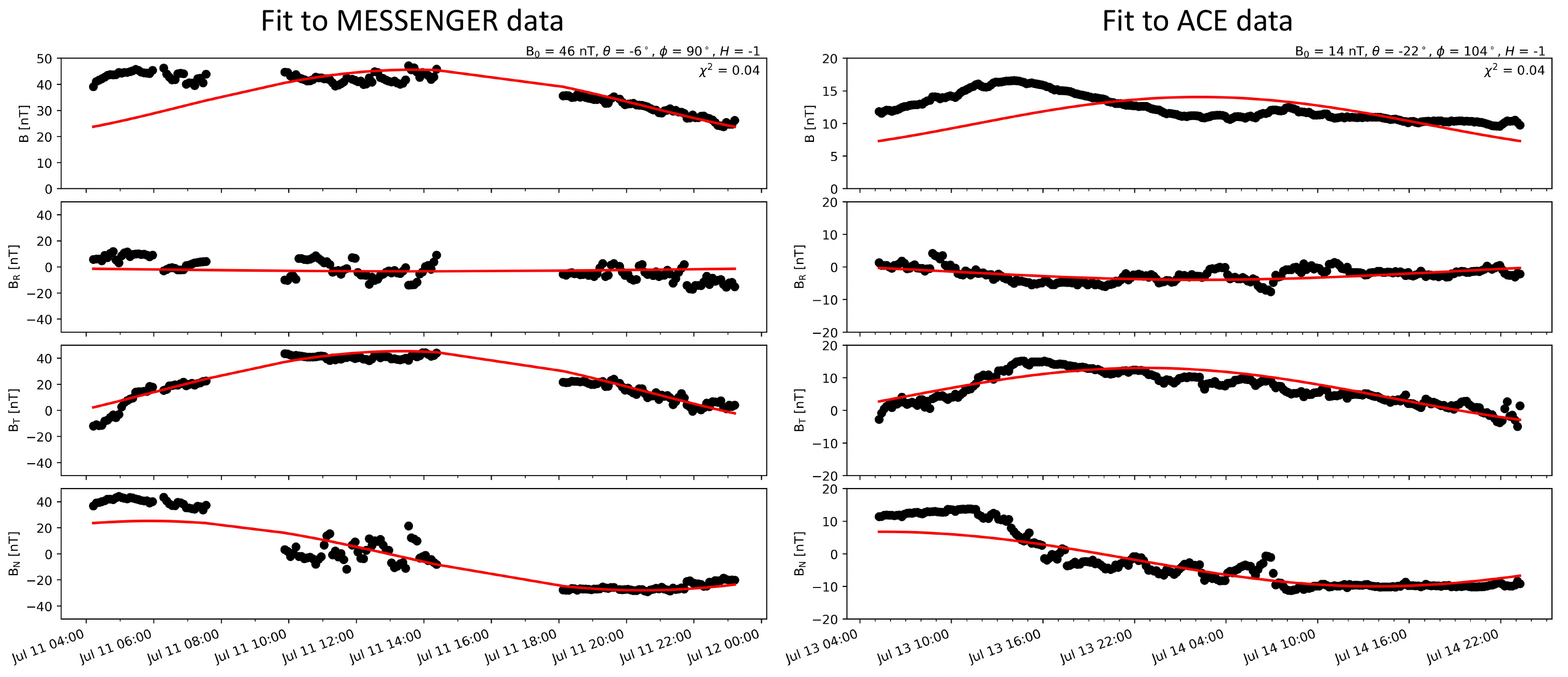}}
\caption{Linear constant $\alpha$ force-free flux rope fits to RTN magnetic field data for the July 9, 2013 CME event at MESSENGER (left column) and ACE (right column).} 
\label{fig:event1_fits}
\end{figure}

The force-free field fits at both spacecraft (Figure~\ref{fig:event1_fits}) also yield a negative helicity flux rope at both spacecraft, with $\theta \sim {-6}^{\circ}$, $\phi \sim 90^{\circ}$ and $B_0 \sim 46$~nT at MESSENGER, and $\theta \sim {-22}^{\circ}$, $\phi \sim 104^{\circ}$, and $B_0 \sim 14$~nT at ACE. Here, $\theta$ is the angle between the flux rope axis and the ecliptic plane, $\phi$ is the angle from the antisunward direction anticlockwise to the projection of the axis direction onto the ecliptic plane, and $B_0$ is the field strength along the flux rope axis. The impact parameter was small at both spacecraft (0.0{4} at MESSENGER and 0.0{5} at ACE), where the impact parameter is defined as the distance of closest approach of the spacecraft to the flux rope axis normalized by the radius of the flux rope, indicating that both spacecraft likely crossed the ICME near the flux rope axis. It is also worth mentioning that the fits had low $\chi^2$ values of 0.04 at both spacecraft, indicating good quality fits. 
To get an estimate of how close to the center/flanks the spacecraft crossed through the ICME structure, we compute the location angle $\lambda$ as defined by \citet{Janvier2013}, based on the results of force-free fittings at MESSENGER and ACE. This yields $| \lambda | \sim 0^\circ$ at MESSENGER, and $| \lambda | \sim 13^\circ$ at ACE, both indicating spacecraft crossings close to the ICME center.

\begin{figure}
\centering
{\includegraphics[width=\linewidth]{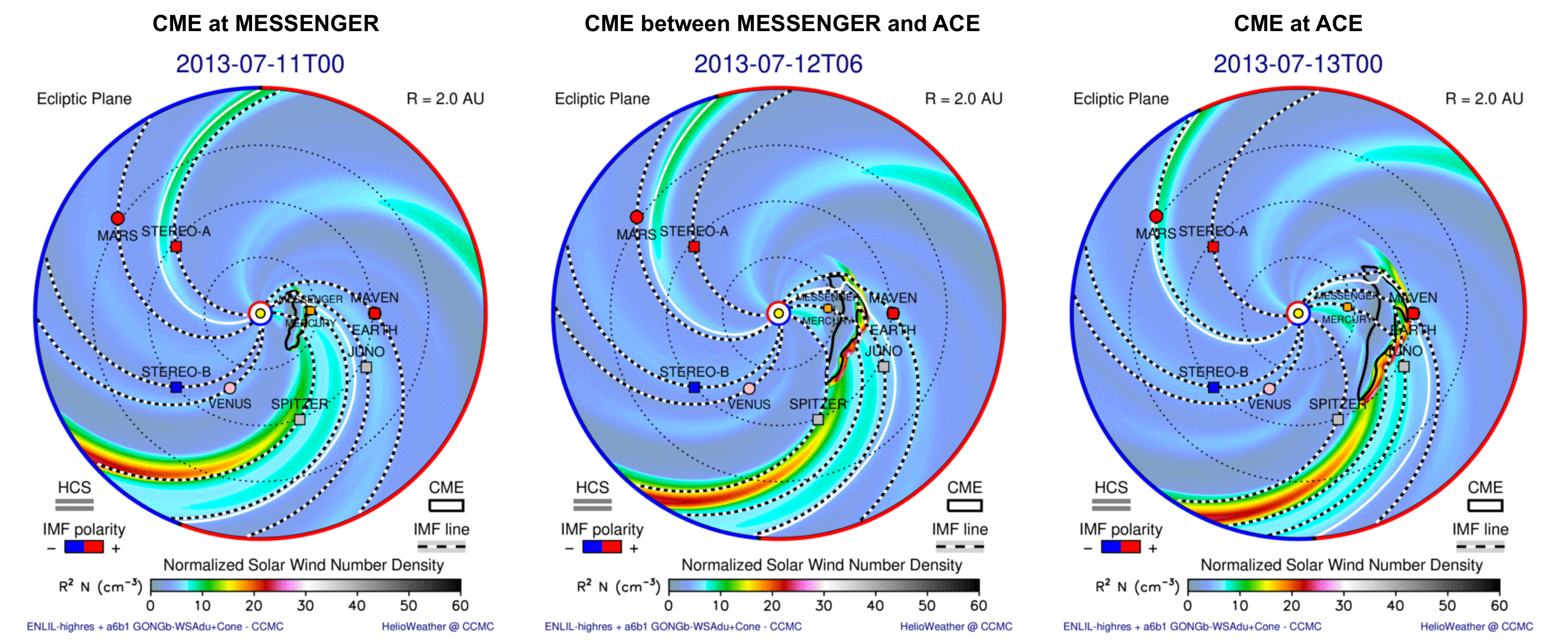}}
\caption{WSA-ENLIL model simulated {heliospheric} conditions for the July 9, 2013 CME event at MESSENGER and ACE, at three time steps: 
(a) at 00~UT on 11 July 2013, just {before} the ICME reached MESSENGER;
(b) at 06~UT on 12 July 2013, during the ICME propagation from MESSENGER to ACE, 
and (c) at 00~UT on 13 July 2013, {just before} the ICME reached ACE. 
{In the simulation, the} HCS reached Mercury {together with the ICME shock front} (a). 
{The two propagated together} all the way to 1~AU (b), 
and eventually reached ACE {at the same time} (c).
{The snapshots have been chosen to be representative of the various phases of the CME propagation in the WSA-ENLIL simulation, and therefore, they can be a few hours off compared to the observed arrival time of the ICME at MESSENGER (Figure~\ref{fig:event1_messenger}) and ACE (Figure~\ref{fig:event1_ace}).}
}
\label{fig:event1_enlil}
\end{figure}

The surprisingly little change in the magnetic field components of this ME from Mercury to Earth prompted us to further investigate the conditions that allowed for this to happen. We therefore {performed a} simulation of the inner heliosphere as described in Section~\ref{sec:methods} {(\url{https://ccmc.gsfc.nasa.gov/database_SH/Camilla_Scolini_010421_SH_1.php}), using the WSA-ENLIL model available at the NASA CCMC to provide context for the background solar wind and global CME propagation.
In the simulation, we initialized the CME using the cone model and parameters from the DONKI catalog.}
Figure~\ref{fig:event1_enlil} shows the modelled solar wind conditions in the {ecliptic} plane during the ICME propagation from Mercury to Earth.

As discussed above, in situ observations indicate that the ICME hit Mercury early on July 11 and arrived at ACE in the second half of day on July 12. Furthermore, WSA-ENLIL simulations indicate that the HCS arrived at Mercury around the same time as the ICME did (left panel in Figure~\ref{fig:event1_enlil}). Therefore, any interaction between the ICME and the HCS occurred at or before the orbit of Mercury. As the ICME traveled from Mercury to ACE faster than the HCS (as visible from the rightmost panel in Figure~\ref{fig:event1_enlil}, which depicts the {interplanetary space} conditions around the time the ICME arrived at ACE), there was likely no structure (e.g. SIRs/CIRs, HCS) in the propagation space of the ICME as it traveled from Mercury to ACE. 
Magnetic field data supports this at both MESSENGER and ACE: at MESSENGER, $B_R$ changed sign (from positive to negative) on July 11, just as the ICME arrived, while the HCS arrived at ACE after the ICME had gone by, likely on or after July 17.
{We note that the CME arrival time at MESSENGER in the WSA-ENLIL simulation is very consistent with in situ observations at MESSENGER (less than 1 hour difference), while the arrival time at ACE is modelled to occur about 7~hours earlier than observed (which is still within the typical uncertainties associated with CME arrival time predictions in such a models, \citet{Riley2018}). 
For this reason, the time steps chosen in Figure~\ref{fig:event1_enlil} are not exactly matching the observed arrival times of the ICME at MESSENGER (Figure~\ref{fig:event1_messenger}) and ACE (Figure~\ref{fig:event1_ace}).}

\section{CME with corotating structures in the propagation space}
\label{sec:event2}

The second CME under study was launched from the Sun on 26 December 2013 at 03:24~UT, and it was later observed in situ in conjunction between MESSENGER and STEREO-B, even though there was a fairly large longitudinal separation (of $32^{\circ}$) between the two spacecraft at that time. At the time of the ICME impact, the MESSENGER spacecraft was orbiting Mercury at HEEQ coordinates $(r,\theta,\phi) = (0.46~\mathrm{AU}, -3^\circ, 177^\circ)$, while STEREO-B was at HEEQ coordinates $(r,\theta,\phi) = (1.09~\mathrm{AU}, 5^\circ, -151^\circ)$. 
\citet{Salman2020} list this CME as a conjunction event, and {predictions of the arrival of the event at various locations based on WSA-ENLIL simulations of the CME listed on the DONKI catalog} also {indicate} the ICME hit both Mercury and STEREO-B as the result of its very large half angular width ($\sim90^{\circ}$ based on the WSA-ENLIL model input available on DONKI). 
{This result is confirmed in the WSA-ENLIL simulation of the CME event performed in this work, as later discussed in Subsection~\ref{subsec:event2_43} and shown in Figure~\ref{fig:event2_enlil}.}



\subsection{Differences in in situ data at the two spacecraft}

\begin{figure}
\centering
\includegraphics[width=\linewidth]{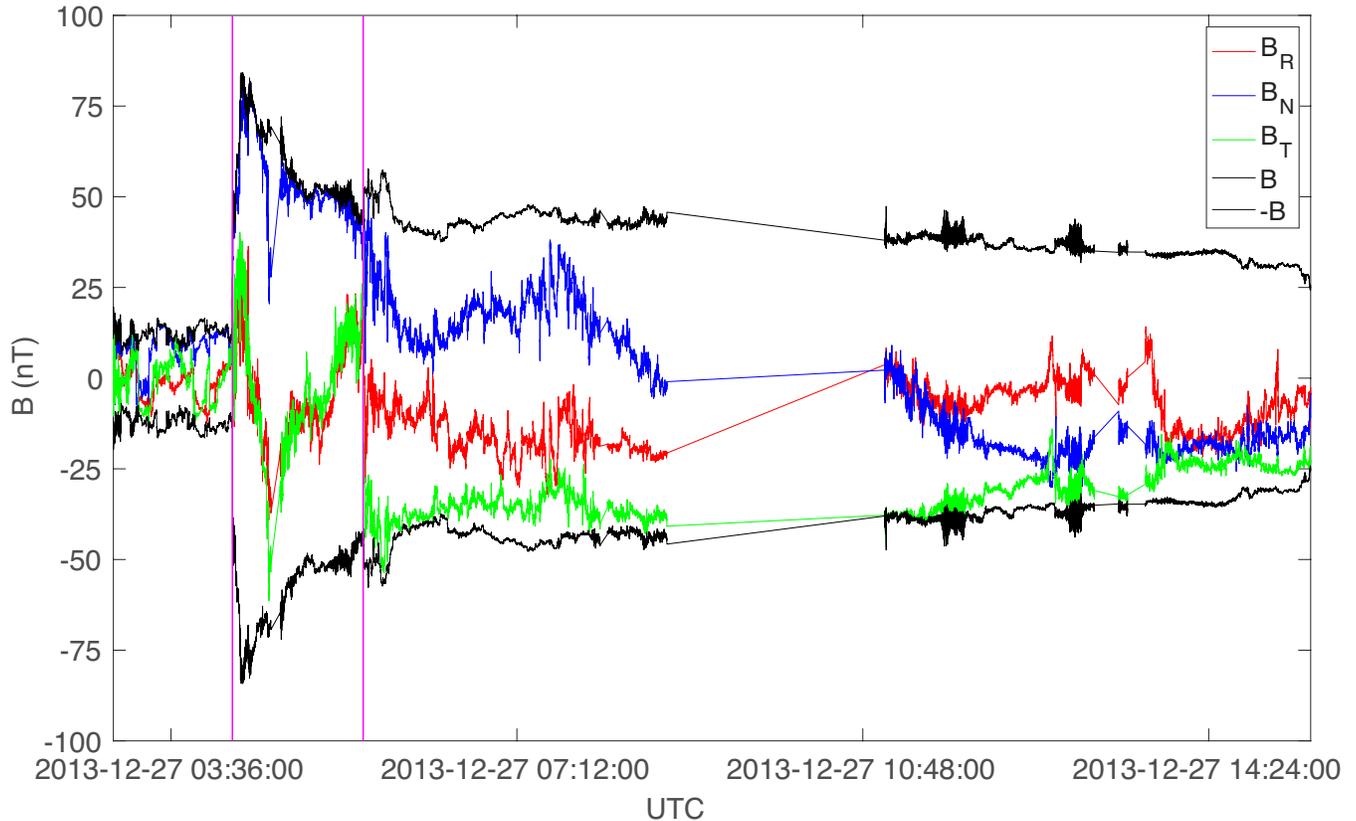} 
\caption{
Magnetic field data for the December 26, 2013 CME event measured in situ at MESSENGER on December 27, 2013. The colors are the same as in Figure~\ref{fig:event1_messenger}.
The vertical magenta lines denote the start time of the shock and ME, while the data interval shown ends at the end of the ME just after which MESSENGER re-enters Mercury's magnetosphere. The data gap corresponds to MESSENGER's passage through the magnetosphere. 
} 
\label{fig:event2_messenger}
\end{figure}

From the catalogue of \citet{Salman2020}, the ICME shock arrived at MESSENGER on December 27 (day of year 361) at 04:14~UT, while the magnetic ejecta arrived at 05:36~UT and ended at 15:28~UT on the same day. This implies a sheath length of $\sim 1.25$ hours and magnetic ejecta length of $\sim 10$ hours at MESSENGER, corresponding to a sheath-to-ME duration ratio equal to 0.13.
Figure~\ref{fig:event2_messenger} shows the magnetic field magnitude and components in RTN coordinates at MESSENGER, with the magnetospheric pass of Mercury removed from the observations. The magnetic field data at Mercury show a fairly simple ICME, with the magnetic ejecta dominated by the $B_T$ magnetic field component and a clear rotation visible in the $B_N$ component.

\begin{figure}
\centering
{\includegraphics[scale=0.7]{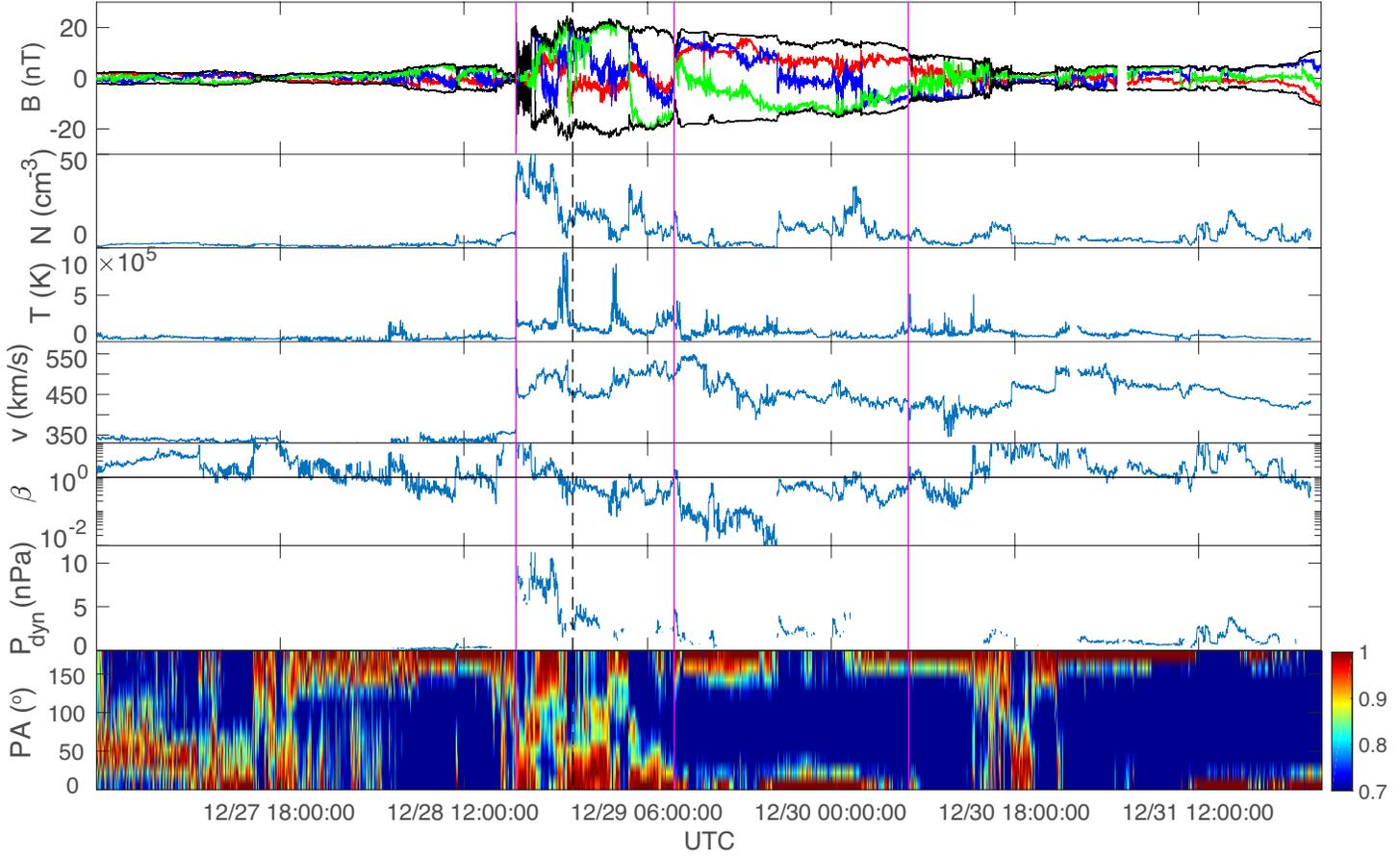}}
\caption{
Data for the December 26, 2013 CME event measured in situ December 28-30, 2013 at STEREO-B. The first panel shows the magnetic field data in RTN coordinates, using the same colors as in Figure~\ref{fig:event1_messenger}.
The second, third, and fourth panels show the proton number density, temperature, and speed, respectively.
The fifth and sixth panels show the solar wind dynamic pressure and plasma $\beta$.
The last panel shows the normalized suprathermal electron pitch angle distribution. The solid vertical magenta lines denote the arrival time of the shock and the start and end times of the actual ME, while the dashed vertical black line marks the start of an additional hypothetical ME that we rule out in Section 4.1 above.
}
\label{fig:event2_stereob}
\end{figure}

At STEREO-B, \citet{Salman2020} list the ICME shock arrival on December 28 at 17:06~UT, the ejecta arrival on December 29 at 04:12~UT, and the ejecta end on December 30 at 14:00~UT. Figure~\ref{fig:event2_stereob} shows the magnetic field magnitude and components (also in RTN coordinates), the solar wind density, temperature, speed, plasma $\beta$, dynamic pressure, and the suprathermal electron pitch angle distribution at STEREO-B. From a quick glance at this data, it is clear that the ICME observed at STEREO-B is more complex than what had been observed at MESSENGER. 

First, the beginning and end boundaries of the magnetic ejecta are not as simple to identify at STEREO-B as they are at MESSENGER. Because we want to explore the changes in the magnetic ejecta during propagation in detail, it is important first to properly identify the boundaries of the ME. Due to the long duration of the ICME and the number of rotations in the magnetic field, we considered the possibility that there could be two magnetic ejecta at STEREO-B; one ME between December 28 22:40~UT and December 29 08:35~UT (marked by the dashed and solid vertical lines on Figure~\ref{fig:event2_stereob}), while another one between December 29 08:35~UT and December 30 07:33~UT (marked by the two solid vertical lines in Figure~\ref{fig:event2_stereob}). However, we can rule out this possibility based on the fact that the first ``ME" does not exhibit typical ME signatures. In this region there is: 
1) a high variability in the $B_N$ component and no smooth rotation in any of the field components; 
2) high solar wind density and temperature;
3) plasma $\beta$ $\sim$1; and 
4) highly variable suprathermal electron pitch angle distribution, with a clear lack of bidirectional electrons. 
Furthermore, and perhaps more importantly, as described in more detail below, there is no other ICME in the propagation space that this ICME is likely to have interacted with and that could have led to the observation of a double magnetic ejecta at STEREO-B but not at MESSENGER.

Based on these considerations, we chose the magnetic ejecta start time to be December 29 08:35~UT, which coincides with the start of the smooth rotation in the magnetic field, drops in solar wind density, temperature, and plasma $\beta$, as well as less variable suprathermal electron pitch angle distribution. We chose December 30 07:33~UT as the end of the ME in order to be able to conduct flux rope fitting; however, we note that the ICME likely continued on until December 30 14:00~UT as indicated by the increased magnetic field in this region. This $\sim7$ hrs long ``back" region of the ICME could have formed through reconnection with the interplanetary magnetic field as suggested by \citet{Dasso2007}.  
These boundary choices then lead to a sheath length of $\sim15.5$ hrs and an ME length of $\sim23$ hrs, 
which correspond to a sheath-to-ME duration ratio equal to 0.67.
By considering the sheath-to-ME duration ratios at the two spacecraft, we note that contrarily to the first ICME case study considered in Section~\ref{sec:event1}, in this case a significantly larger ratio is observed at 1~AU (0.67) compared to Mercury (0.13). This indicates a much more extended sheath relative to the ejecta duration at 1~AU, than was observed at Mercury. Furthermore, based on the boundary choices at the two spacecraft, we recover an ME increase factor equal to 2.3, i.e.\ very similar to the typical ME increase factor of 2.6 observed between these two heliocentric distances \citep{Janvier2019}. On the other hand, the increase in the sheath duration is significantly larger, around 12.4, compared to a typical increase factor of only 5.
Similar evidence of an over-increase of the sheath duration compared to typical can be recovered based on the relations proposed by \citet{Salman2020}, who reported a typical sheath duration increase with heliocentric distance as $\Delta t_{sh} = 15.28 \cdot r - 3.67$ (in hrs). Using this relation to estimate the expected sheath duration at MESSENGER and STEREO-B distances, one would have expected an increase of a factor 3.9 between the two spacecraft, which is significantly lower than the observed one (12.4).

In order to estimate the expected sheath and ME growth from expansion only, we fit the typical sheath and ME duration provided by \citet{Janvier2019} at selected heliocentric distances in the inner heliosphere with a power law function. 
We recover exponents equal to $1.77$ and $1.08$, respectively, which we take as indicators of the typical sheath and ME growth rates between Mercury and 1~AU.
By extrapolating the structures' size from MESSENGER to STEREO-B assuming the typical growth rates above, 
we estimate the following duration at 1~AU: 
$\Delta t_{sh} = 1.25 \cdot (1.09/0.46)^{1.77}$~hrs~$\sim 5.8$~hrs for the sheath, and
$\Delta t_{ME} = 10 \cdot (1.09/0.46)^{1.08}$~hrs~$\sim 25.4$~hrs for the ME.
For comparison, the 23-hour long ME observed at STEREO-B well matched the expected size estimated from a typical increase rate.
On the other hand, the observed sheath duration is a factor 2.7 larger than predicted using empirical relations, which again highlights the anomalous growth of this structure during propagation between Mercury and 1~AU.

We can also compare the dynamic pressure observed at STEREO-B to that at MESSENGER. As described in Section~\ref{sec:methods}, 
we estimate the dynamic pressure ($P_{dyn}$) from the outbound portion of MESSENGER's magnetospheric pass around Mercury, which occurred during the middle of the magnetic ejecta of this ICME event. This yields $P_{dyn} \sim 5.5$~nPa, which is close to the average value of 7~nPa at Mercury's heliocentric distances \citep{Winslow2013}. Although this is a single point estimate, which does not allow us to put bounds on the maximum dynamic pressure that could have been observed by MESSENGER during this time, we note that Mercury's magnetosphere was not highly disturbed by this ICME event. This is another indication that the dynamic pressure was likely not unusually high during this time, as events with very high $P_{dyn}$ are known to cause extreme changes in Mercury's magnetosphere (e.g., magnetopause compressed to the surface, very large cusps, etc.; e.g. \citet{Winslow2020}). At STEREO-B, the maximum observed $P_{dyn}$ is 11.3~nPa, and the average value in the ICME is 2.4~nPa. Scaling this to Mercury's location with a $1/r^2$ scaling yields a maximum $P_{dyn} \sim 63$~nPa and an average value of 13~nPa, much higher than the value estimated at MESSENGER. Thus the fact that the observations indicate a fairly average dynamic pressure event at MESSENGER while at STEREO-B we observe a high dynamic pressure event provides further evidence that the ICME has sustained significant changes in propagation between MESSENGER and STEREO-B. Furthermore, to explain such a high dynamic pressure observed at STEREO-B compared to what was estimated at MESSENGER, we expect the ICME to have interacted with a dense structure, such as a SIR/CIR in the intervening interplanetary space (further justification in support of this hypothesis is given in Section~\ref{subsec:event2_43} below).

\begin{figure}
\centering
{\includegraphics[width=\linewidth]{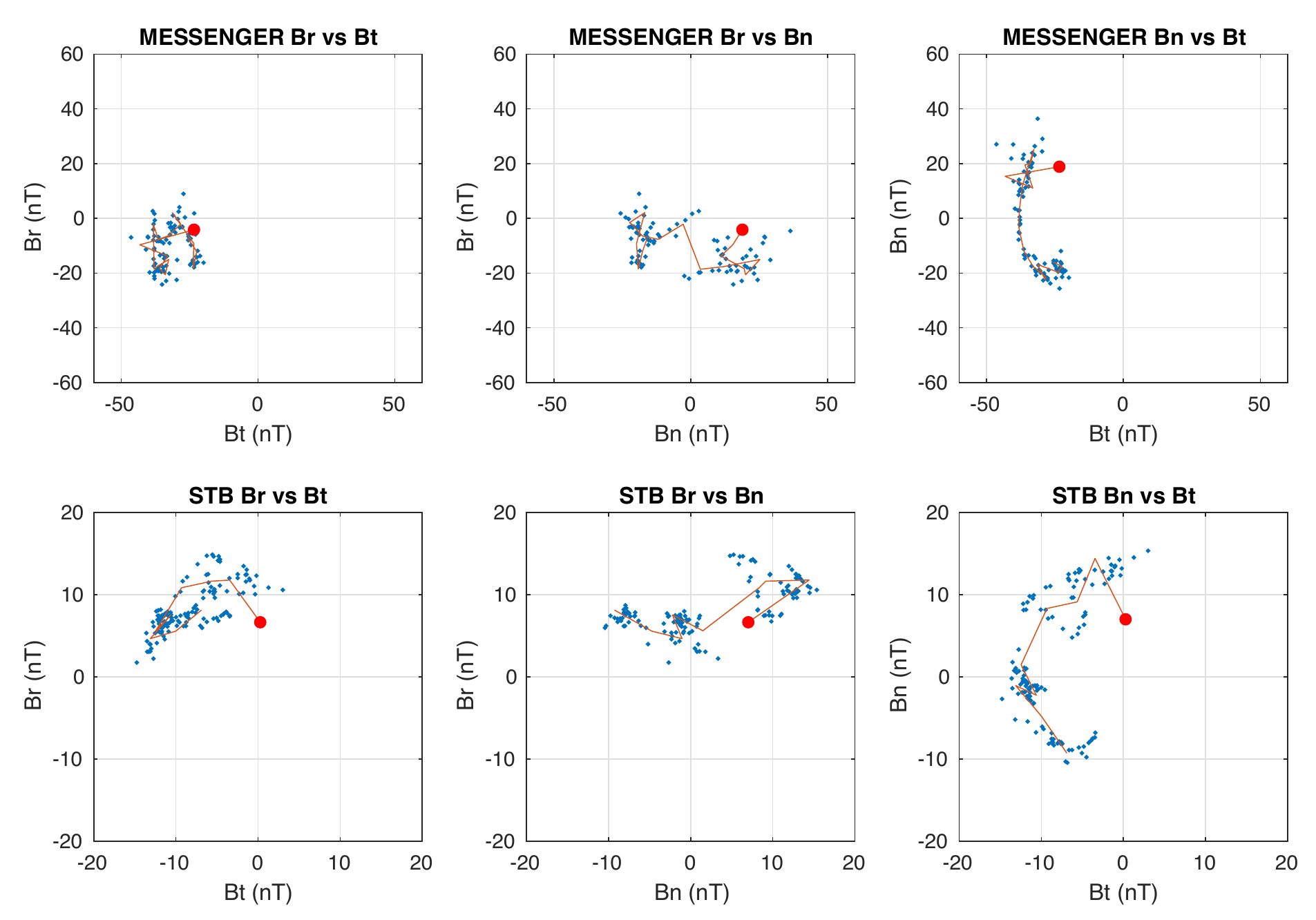}}
\caption{
Magnetic hodograms of the December 26, 2013 CME event at MESSENGER and STEREO-B, in RTN coordinates. The red dots mark the initial values of the magnetic field components.}
\label{fig:event2_hodograms}
\end{figure}

Using the classification of \citet{Nieves2019} and the hodograms in Figure~\ref{fig:event2_hodograms}, 
we find an $F^-$ flux rope type at MESSENGER, and an $Fr$ flux rope type at STEREO-B. Looking at the $B_N$ and $B_T$ components only in Figure~\ref{fig:event2_hodograms} (right-hand top and bottom panels) imply a NES type flux rope at both spacecraft, with the lack of a full $180^{\circ}$ rotation at MESSENGER possibly due to the spacecraft crossing far from the flux rope axis. However, it is clear from the left and center panels of Figure~\ref{fig:event2_hodograms} that there is a significant $B_R$ component of opposite sign at the two spacecraft, which complicates the classification. This could be due to one spacecraft crossing far above and the other spacecraft crossing far below the flux rope axis, but we also note that in situ flux rope observations that have a significant $B_R$ component can also be due to a flux rope with a substantial tilt in longitude ($\phi$ direction). However, since the initial direction of the CME in latitude was $\theta\sim-15^{\circ}$ (see next section for details), in this case it is likely that the flux rope axis passed south of both MESSENGER and STEREO-B (based on the spacecraft's latitude at this time), and thus that both spacecraft crossed the ICME above the flux rope axis, which should have yielded positive $B_R$ components at both spacecraft. Based on the fact that opposite sign $B_R$ components are observed at MESSENGER and STEREO-B, we infer that the most likely reason for the significant $B_R$ differences at the two spacecraft is due to the flux ropes having undergone a substantial rotation in the $\phi$ direction with respect to the $\phi=90^{\circ}$ line, as further discussed below and in Section~\ref{subsec:event2_42}.

\begin{figure}
\centering
{\includegraphics[width=\linewidth]{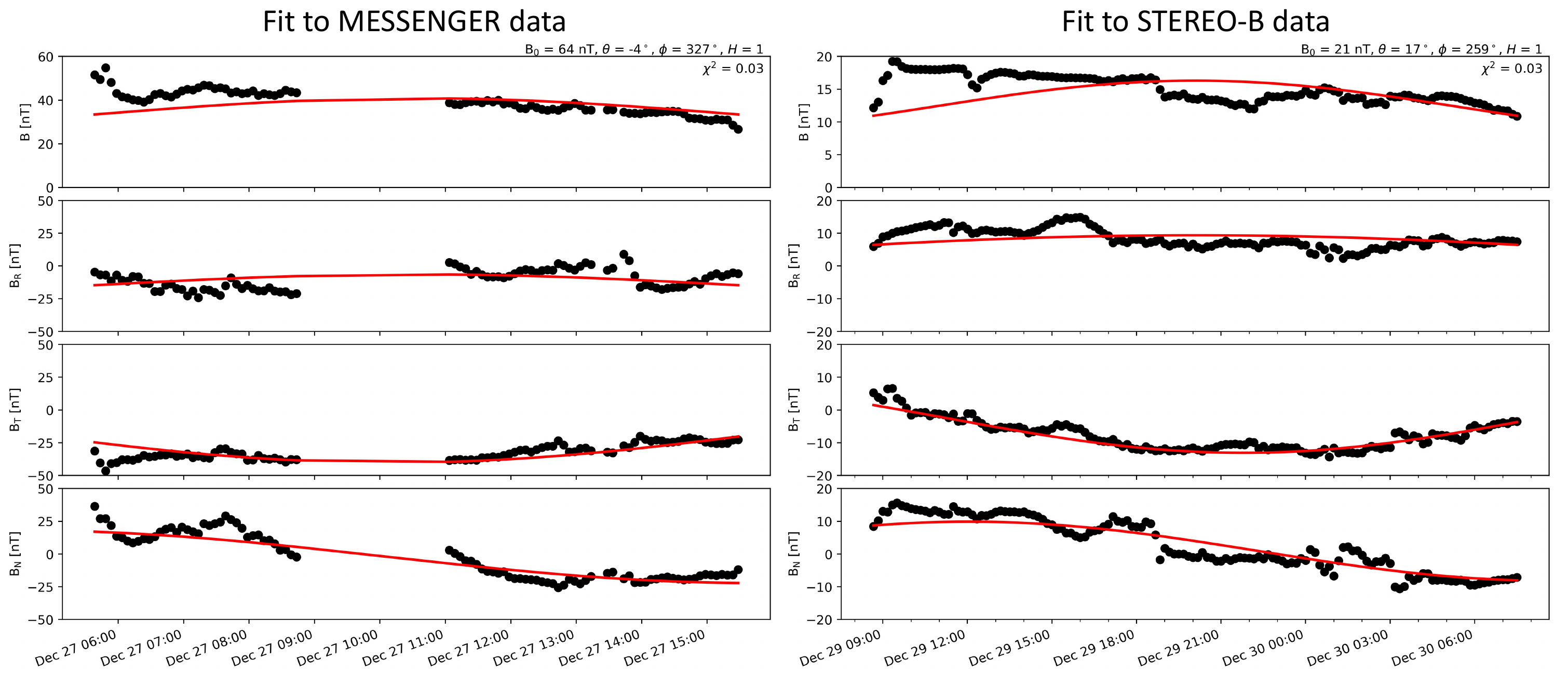}} 
\caption{
Linear constant $\alpha$ force-free flux rope fits to RTN magnetic field data for the December 26, 2013 CME event at MESSENGER (left column) and STEREO-B (right column).}
\label{fig:event2_fits}
\end{figure}

The force-free field fits (Figure~\ref{fig:event2_fits}) yield a positive helicity flux rope at MESSENGER, with $\theta\sim-4^{\circ}$, $\phi\sim327^{\circ}$ and $B_0\sim64$~nT, and a {positive} helicity flux rope at STEREO-B, with $\theta \sim {17}^{\circ}$, $\phi \sim {259}^{\circ}$, and $B_0\sim21$~nT. The impact parameter was large at both spacecraft, with a value of 0.{6} at MESSENGER and 0.{8} at STEREO-B, and the fits had low $\chi^2$ values of {0.03 at both spacecraft.} {The $\sim$70$^{\circ}$ rotation in $\phi$ between MESSENGER and STEREO-B estimated from the flux rope fits is in agreement with the $B_R$ changes noted in the hodograms, and underlines the likelihood that the $B_R$ difference between the two spacecraft are caused by rotation of the flux rope and not due to the spacecraft crossing location with respect to the flux rope axis. This provides an additional indication that the ICME structure underwent significant changes during propagation, and that those changes involved an increase in complexity in the flux rope structure. Evidence of a complex structure at 1~AU is also given by suprathermal electron data at STEREO-B, which shows only very short sections of bidirectional electrons inside the ME. Such characteristics are compatible with extended magnetic reconnection and a global alteration of the flux rope magnetic connectivity. We also note that at the Sun, EUVI-B observations of the CME source region hint towards the eruption of a right-handed flux rope {(as indicated by the forward-S topology of pre-eruptive coronal loops in Figure~\ref{fig:event2_remote_sensing}a)}, although the lack of photospheric magnetic field measurements prevents an accurate estimation of the source region and flux rope magnetic properties, such as the helicity and initial orientation/flux rope type. Overall, the eruption of a right-handed flux rope is consistent with in situ estimates of a positive helicity flux rope at MESSENGER and STEREO-B as obtained from the force-free field fits.}

\subsection{Ruling out possible factors affecting change in this ICME}
\label{subsec:event2_42}

In the above section, we presented significant differences in the ICME sheath duration, dynamic pressure, and flux rope orientation observed in situ between MESSENGER and STEREO-B. As discussed below, we can rule out two possible causes of these differences: 1) due to ICME flank arrival at MESSENGER vs center arrival at STEREO-B, and 2) due to interaction with other ICMEs in the propagation space.

\begin{figure}
\centering
{\includegraphics[width=\linewidth]{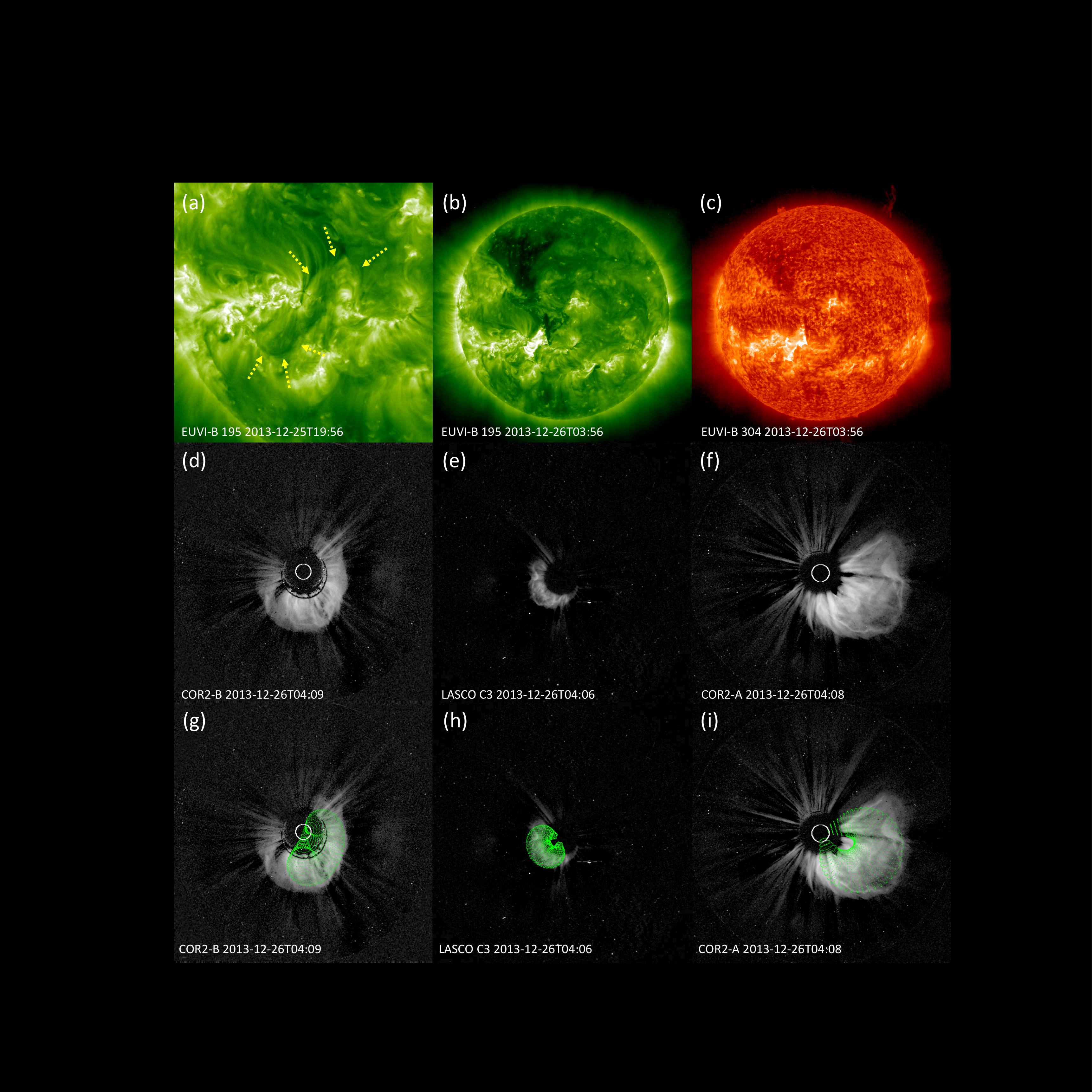}} 
\caption{
{
Overview of the remote-sensing observations of the December 26, 2013 CME. 
(a) EUVI-A image of the CME source region in the 195\AA \ filter on December 25, 19:25~UT, before the eruption. The skew of some coronal loops, marked by the yellow dotted lines, suggests a positive helicity region. 
(b) EUVI-A full disk image in the 195\AA \ filter on December 26, 03:56~UT, after the eruption.  
(c) EUVI-A full disk image in the 304\AA \ filter on December 26, 03:56~UT, after the eruption. 
Bright post eruptive arcades (b) and flare ribbons (c) are clearly visible in the source region.
(d) -- (i): base-difference white-light coronal images of the CME as seen by STEREO-B COR2 (left), SOHO LASCO C3 (middle), and STEREO-A COR2 (right) around December 26, 04:08~UT. The CME is visible as a bright feature. Panels (d), (e), (f) show the plain images, while in panels (g), (h), (i) the GCS wireframe is overplotted (in green).}
}
\label{fig:event2_remote_sensing}
\end{figure}

{To discuss point 1) above, we consider remote-sensing observations, along with the location angle derived from the flux rope fits, duration of the ICME, and WSA-ENLIL model simulations. An overview of the remote-sensing observations of the CME and its source region is provided in Figure~\ref{fig:event2_remote_sensing}.}
Based on STEREO EUVI-B images {(Figure~\ref{fig:event2_remote_sensing}, a--c)}, the CME under study erupted around 02:30~UT on December 26 from a source region located around $\theta\sim -11^{\circ}$ and $\phi\sim -165^{\circ}$ in Stonyhurst coordinates. The CME was later observed in the corona by the LASCO~C2 and C3 instruments (starting at 03:12~UT and 03:48~UT, respectively), as well as by the COR2 instruments on board STEREO-A and STEREO-B (first observations at 03:24~UT and 03:55~UT, respectively). In the corona, the CME initial direction (as listed in the DONKI database) was towards $-134^{\circ}$ HEEQ longitude, and at this time STEREO-B was at $-151^{\circ}$ HEEQ and MESSENGER at $176^{\circ}$ HEEQ. Thus, the CME most probably underwent a deflection of about $30^\circ$ westward within the first solar radii of propagation. We conducted a reconstruction of the CME geometry and kinematics with the GCS model and multiple viewpoints (i.e.\ observations from LASCO, STEREO-A and STEREO-B{ (Figure~\ref{fig:event2_remote_sensing}, d--i)}, which indicates that the CME in the corona was directed towards $-15^{\circ}$ HEEQ latitude and $-135^{\circ}$ HEEQ longitude, essentially confirming the DONKI-based longitude.

{Thus, remote-sensing observations suggest the CME initial direction was more closely aimed towards STEREO-B than towards Mercury, raising the possibility that MESSENGER passed through the flanks of the ICME while STEREO-B passed through closer to the center, which could partly explain the larger differences between the observations at the two locations. However, from in situ observations, for a fully flank arrival we would expect a $B_R$ dominant field with very little rotation in the magnetic field direction overall \citep{Moestl2010, Moestl2020}, and a significantly longer duration ICME crossing than for a center arrival. In fact at MESSENGER we see a $B_T$ dominant field, with clear rotation in $B_N$, and a much shorter duration ICME crossing (even after accounting for expansion) than at STEREO-B. Based on these arguments, we can conclude that the ICME likely had a close to center arrival at STEREO-B and an in-between flank and center arrival at MESSENGER. This picture is further supported by the location angle $\lambda$ calculated at MESSENGER and STEREO-B based on results from the force-free fittings. We obtain $| \lambda | \sim 57^\circ$ at MESSENGER, and $| \lambda | \sim {10}^\circ$ at STEREO-B, suggesting that MESSENGER passed through closer to the flank of the ICME than STEREO-B, but not actually in the flank (which would be at $|\lambda | \sim 90^\circ$). This is in line with WSA-ENLIL simulation results shown in Figure~\ref{fig:event2_enlil}. Overall, although the two spacecraft did not cross through the exact same portion of the ICME, it is also clear that MESSENGER did not cross through the flanks and the observational differences are too large to be explained by crossing location alone.}



Regarding point 2) from above, coronal and heliospheric remote-sensing observations in the days before and after the CME eruption show little activity, and thus it is not likely that there was any other ICMEs in the propagation space that the ICME in question would have interacted with on its way to MESSENGER and STEREO-B. From STEREO COR2-A and -B, the only other CME around this time (a small CME launched around 01:25~UT on December 25, i.e., the day before the CME in question) was heading towards STEREO-A, and not STEREO-B. MESSENGER was directly in between STEREO-A and STEREO-B at this time, which implies that for this CME to have been in the propagation space of the ICME in question between MESSENGER and STEREO-B, this smaller, earlier event would have had to have been observed at MESSENGER. As there were no ICMEs observed at MESSENGER in the days prior to December 26, we can rule out this small CME having interacted with and caused the observed changes in the CME studied here by the time in reached STEREO-B.

\subsection{Critical factor affecting this ICME: SIR}
\label{subsec:event2_43}

Although the $\sim30^{\circ}$ separation between the observing spacecraft could inherently introduce differences in the measured magnetic field, the differences observed (both in terms of sheath length, dynamic pressure, and magnetic field structure and complexity) are too large to be explained by just the separation alone. Similarly to the event showcased in \citet{Winslow2016}, the ICME magnetic field is fairly simple at MESSENGER, but it is much more complex by the time it reaches STEREO-B.

\begin{figure}
\centering
{\includegraphics[width=\linewidth]{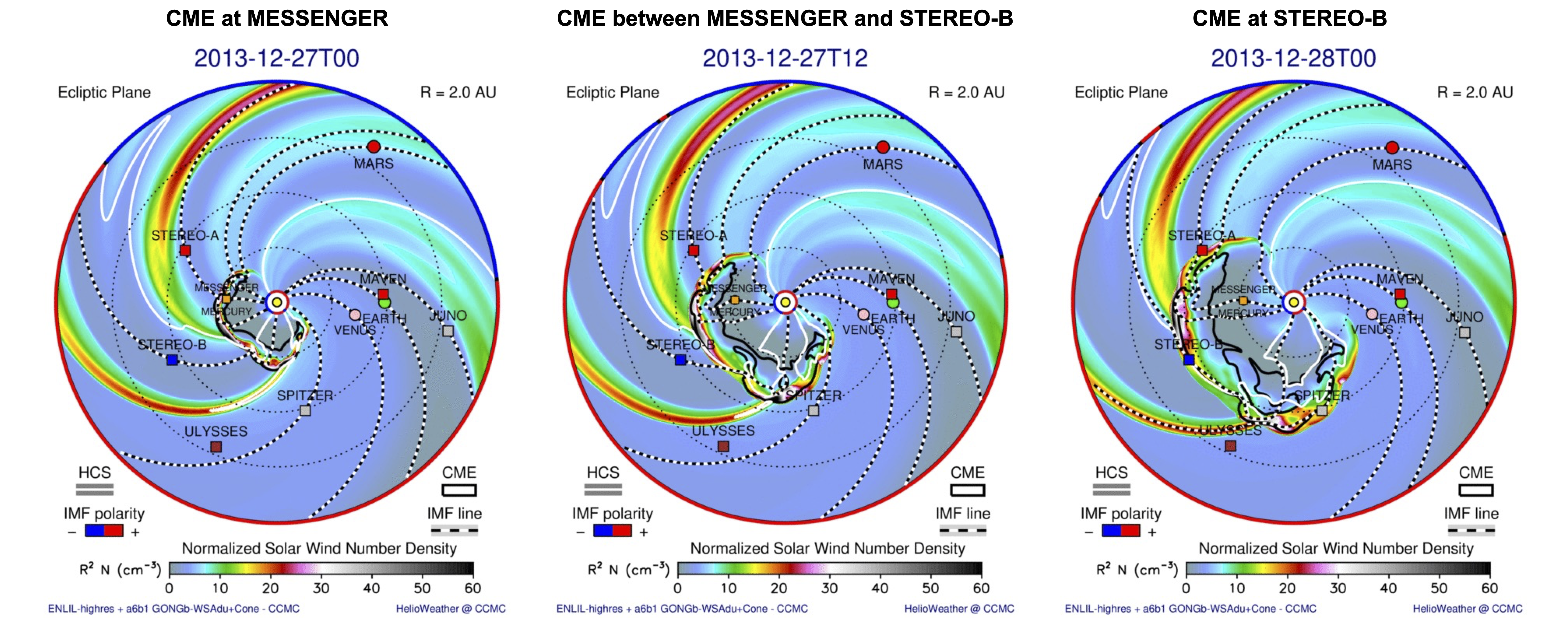}} 
\caption{
WSA-ENLIL model simulated {heliospheric} conditions for the December 26, 2013 CME event at MESSENGER and STEREO-B, at three time steps: 
(a) at {00}~UT on {27} December 2013, just after the ICME reached MESSENGER;
(b) at {12}~UT on {27} December 2013, during the ICME propagation from MESSENGER to STEREO-B, 
and (c) at {00}~UT on {28} December 2013, just after the ICME reached STEREO-B. 
{In the simulation, the} HCS and SIR had reached Mercury prior to the ICME arrival (a),
propagated in the ICME propagation space between MESSENGER and 1~AU (b), and eventually reached STEREO-B after the ICME had passed by (c).
{We note that the snapshots have been chosen to illustrate the interaction between the CME and the SIR/HCS in the WSA-ENLIL simulation, and therefore, they do not necessarily match the observed arrival time of the ICME at MESSENGER (Figure~\ref{fig:event2_messenger}) and STEREO-B (Figure~\ref{fig:event2_stereob}).}
} 
\label{fig:event2_enlil}
\end{figure}

We therefore use WSA-ENLIL data as described in Section~\ref{sec:methods} to contextualize the plasma environment through which the ICME propagated between Mercury and STEREO-B.
WSA-ENLIL simulations (available at: \url{https://ccmc.gsfc.nasa.gov/database_SH/Camilla_Scolini_010421_SH_2.php}) of the background  solar wind {and of the CME initialized with parameters from the DONKI catalog} (Figure~\ref{fig:event2_enlil}) show that there was an SIR, including the HCS, in the ICME's propagation space from Mercury to STEREO-B at this time. 
{In this case, the CME arrival time at MESSENGER ad STEREO-B was modelled by ENLIL to occur about 7 and 20~hours earlier than observed, respectively. In this case, the error on the arrival time is therefore significantly larger than for the first CME event considered, and for this reason, the time steps chosen in Figure~\ref{fig:event2_enlil} are not matching the observed arrival times of the ICME at MESSENGER (Figure~\ref{fig:event2_messenger}) and STEREO-B (Figure~\ref{fig:event2_stereob}). 
Despite this discrepancy, key factors affecting the CME propagation and interaction with the ambient solar wind can be identified in Figure~\ref{fig:event2_enlil}, and are further supported by simulations of the steady-state solar wind alone (available at: \url{https://ccmc.gsfc.nasa.gov/database_SH/Dan_Aksim_060419_SH_4.php}, not shown here).}
{In particular, the} model predicts the SIR/HCS to have passed by Mercury by the time the ICME arrives there {(left panel in Figure~\ref{fig:event2_enlil})}, while it predicts the SIR/HCS arrival at STEREO-B early on December 29 soon after the ICME arrival {(right panel in Figure~\ref{fig:event2_enlil})}. We note that since here we are only interested in the time period of a few days between the CME eruption time and its arrival at 1~AU, in the following we use the term ``SIR'' as we have no information about whether this specific stream interaction region was also a CIR, i.e. a corotating stream interaction region originated by a recurrent coronal hole and observed in situ over multiple solar rotations \citep{Richardson2018}.

As ENLIL finds that the HCS arrives during the ICME passage at STEREO-B, and as differences of up to a couple of days have been reported between measured HCS crossings and simulated ones \citep[e.g.][]{Lee2009, Szabo2020}, we turn our attention to the in situ data at MESSENGER and STEREO-B.
From magnetic field data at Mercury, we find that the HCS crossing occurred sometime between December 22 and December 24, as MESSENGER was skimming the current sheet and switched from $+B_R$ IMF to $-B_R$ IMF in that time (in agreement with results from the WSA-ENLIL simulations). Thus by the time the ICME reached Mercury on December 27, the HCS was in the propagation space between Mercury and STEREO-B as predicted by the WSA-ENLIL model. At STEREO-B, in situ plasma and magnetic field data suggest that the HCS arrived later in the day on December 27 just before the ICME, as can be seen from the change in direction of the strahl electrons and the sign change in $B_R$ around 18:00~UT on December 27 in Figure~\ref{fig:event2_stereob}. Thus it is clear from in situ data that the HCS arrived at both Mercury and STEREO-B before the ICME, and therefore it is unlikely that the ICME interacted with it. {The WSA-ENLIL model results are in agreement with this conclusion, also showing that the HCS was at the front of the SIR.} It is important to mention that for this second case study, a sub-optimal performance of the WSA-ENLIL model prediction of the solar wind structure in the region of interest is likely given that the photospheric magnetic field data used as input for such simulations are only updated in regions on the solar disk that are seen as front-sided from Earth (i.e.\ opposite to the direction of propagation of this ICME).


Given that we only have magnetic field data at MESSENGER and due to the frequent magnetospheric crossings of Mercury, it is not possible to determine when the SIR itself arrived there. 
However, the STEREO SIR database\footnote{Available at: \url{https://stereo-ssc.nascom.nasa.gov/data/ins_data/impact/level3/LanJian_STEREO_SIR_List.txt}} \citep{Jian2019} lists an SIR start time at STEREO-B on 21:47~UT on December 31, ending at 17:38~UT on 2 January 2014, and there is no other SIR identified for more than 6 days before or after the ICME in question. This suggests that the SIR arrived at STEREO-B two days later than in the WSA-ENLIL simulation. It is also important to note that neither MESSENGER nor STEREO-B were exactly in the heliographic equatorial plane, whereas these simulation results are: this could have been an additional cause of these differences in modeled and observed arrival times.

Overall, based on the WSA-ENLIL simulation and the in situ data, we can infer the most likely scenario that the ICME caught up to and overtook the SIR between Mercury and STEREO-B, which is why the ICME arrived earlier than the SIR as seen in situ at STEREO-B. Furthermore, \citet{Salman2020} list the CME initial speed at $\sim1000$~km/s, the transit speed from Sun to Mercury at $\sim770$~km/s, and the transit speed from Mercury to STEREO-B at $\sim710$~km/s. From STEREO-B data, we find that the SIR had a maximum speed of 540~km/s, similar to the maximum in situ speed of the ICME at that location. Given that we know that the ICME had a nearly 200~km/s larger transit speed from Mercury to STEREO-B, we infer that the ICME was likely traveling significantly faster closer to Mercury than near STEREO-B, encountered the SIR and slowed down as it overtook it due to the substantial drag posed by the high density stream. During this encounter, the ICME was significantly transformed from what was observed at Mercury to that observed at STEREO-B. 

 


\section{Summary and Conclusions}
\label{sec:conclusions}

In this work, we have presented a study of two ICMEs observed at Mercury and 1~AU by spacecraft in longitudinal conjunction. Our interest was driven by the following question: what causes the drastic alterations observed in some ICMEs during propagation, and why do other ICMEs remain relatively unchanged? The two ICMEs under study presented major differences in their evolutionary behaviour, with the first one propagating in a relatively self-similar manner, while the second one underwent significant changes in its properties. The in-depth analysis of the ICME magnetic field and plasma properties at Mercury and at 1~AU together with considerations on the size evolution of sheaths and MEs with heliocentric distance, {as well as simulations of the solar wind and the ICME performed with the WSA-ENLIL + cone model}, allowed us to reconstruct the propagation scenario of the two events between Mercury and 1~AU and to identify the critical factors controlling their evolution during propagation.
The main results can be summarized as follows:
\begin{enumerate}
\item 
The first case study (CME launched from the Sun on 9 July 2013) was a relatively simple ICME observed during a period of near-perfect conjunction ($\sim 3^\circ$ of longitudinal separation) exhibiting little change in its global structure between Mercury and 1~AU. This long-duration ICME presented similar signatures at the two spacecraft locations, such as clear boundaries between the sheath and magnetic ejecta, relatively similar sheath-to-ME ratio at both locations, sheath dynamic pressure values following typical scaling laws, clear bidirectional electrons throughout the whole magnetic ejecta at 1~AU, and little change in the fitted flux rope parameters (e.g.\ handedness, orientation, classification) for the magnetic ejecta at ACE compared to MESSENGER. 
\item 
The second event (CME launched from the Sun on 26 December 2013) was a case where significant changes in the ICME properties and structure were observed between Mercury and 1~AU. Although not observed in perfect conjunction, the longitudinal separation between the two spacecraft ($\sim 30^\circ$) was not enough to explain the large differences observed in situ at the two locations, {especially since we were able to rule out a flank hit at both spacecraft.}
At MESSENGER, the ICME presented simple characteristics consistent with typical ICME properties at this heliocentric distance.
At STEREO-B, on the other hand, a much more complex structure was observed:
first, it proved to be complicated to define the boundaries between the sheath and the magnetic ejecta; 
the sheath duration was also significantly more extended than at MESSENGER, both in absolute terms and with respect to the ejecta duration;
the dynamic pressure in the sheath at STEREO-B was also much greater than what could be expected from the application of simple scaling laws.
Finally, bidirectional electrons, a typical proxy used to identify MEs whithin ICMEs, were largely missing at STEREO-B, indicating that sufficient magnetic reconnection occurred between the ME and the surrounding solar wind to completely alter the magnetic connectivity of the ICME structure.
In addition, we reported significant changes in the fitted flux rope parameters {(e.g.\ orientation, classification) for the ME at STEREO-B compared to MESSENGER, indicating a large rotation in the azimuthal direction of the flux rope axis, further underlining that reconnection in the ME took place during propagation.} 
All of this indicated increased complexity of the very nature of the ICME during propagation through interplanetary space. 
\item 
For both events, we could rule out any interaction with other ICME structures in interplanetary space, as indicated by remote-sensing and in situ instrument data.
For the first case study, we were able to exclude the presence of other solar wind transients (e.g. SIRs, HCS) in the ICME propagation space, therefore providing evidence that ICME structures (e.g.\ magnetic flux ropes, sheaths) remain approximately self-similar during propagation if no interaction with other transients in the solar wind occur. 
For the second case study, WSA-ENLIL simulations revealed the presence of an SIR and HCS propagating between Mercury and STEREO-B around the same time of the ICME. From {these simulations,} in situ data, and timing considerations we found that the HCS was ahead of the ICME and it is unlikely that the ICME could have interacted with it between Mercury and 1~AU, however, the SIR was in between Mercury and 1~AU at the right time for the ICME to have caught up to it, interacted with it, and overtaken it due the ICME's faster speed. This interaction with the high density stream is what caused the large increase in dynamic pressure and sheath duration from MESSENGER to STEREO-B, and the substantial rotation in flux rope orientation. Based on these results, we can conclude that the large changes observed in the ICME structure of this event between MESSENGER and STEREO-B were most likely due to the interaction with the SIR, and not due to any other dominant factor.
\end{enumerate}


Ultimately, the analysis of more events will be needed to provide definitive conclusions on all the possible causes and on the frequency of drastic alterations in ICME structures during propagation.
So far, the capability to extend such an in-depth analysis to a larger set of events has been challenging because of the extremely limited number of ICMEs exhibiting clear in situ flux rope signatures and observed at different radial distances by multiple spacecraft in longitudinal conjunction. However, while MESSENGER data have been extensively explored and analysed with respect to ICME conjunction events \citep{Lugaz2020a, Salman2020}, in the near future a statistical analysis over a larger set of events will likely benefit from new data from the Parker Solar Probe \citep{Fox2016}, Solar Orbiter \citep{Mueller2020} and potentially BepiColombo \citep{Benkhoff2010} missions.


\acknowledgments
Support for this work was provided by NSF grant AGS1622352. 
R. M. W. acknowledges support from NASA grants NNX15AW31G and 80NSSC19K0914, and NSF grant AGS1622352, as well as partial support from the NASA STEREO Quadrature grant. 
C. S. acknowledges the NASA Living With a Star Jack Eddy Postdoctoral Fellowship Program, administered by UCAR's Cooperative Programs for the Advancement of Earth System Science (CPAESS) under award no. NNX16AK22G. 
N. L. acknowledges support from NASA grants 80NSSC17K0556, 80NSSC20K0431 and 80NSSC20K0700. 

All the data analyzed in this study are publicly available. MESSENGER data are available on the Planetary Data System ({\url{https://pds.jpl.nasa.gov}}), while STEREO data are available on the Space Physics Data Facility's Coordinated Data Analysis Web ({\url{https://cdaweb.sci.gsfc.nasa.gov}}).
Simulation results have been provided by the Community Coordinated Modeling Center (CCMC) at NASA Goddard Space Flight Center through their public Runs on Request system (\url{http://ccmc.gsfc.nasa.gov}). The CCMC is a multi-agency partnership between NASA, AFMC, AFOSR, AFRL, AFWA, NOAA, NSF and ONR. The ENLIL model was developed by D. Odstr\v{c}il at the University of Colorado at Boulder.

\bibliographystyle{aasjournal}



\end{document}